\documentclass[sort&compress,a4paper]{spie}  
\usepackage[]{graphicx}
\usepackage[colorlinks=true,citecolor=blue,linkcolor=black]{hyperref}
\usepackage{amsmath}
\usepackage{amssymb}
\usepackage{subfigure}
\newcommand{\veq}{\varepsilon}
\newcommand{\csa}{C_\textrm{scat}}

\title{Mueller matrix polarimetry of plasmon resonant silver nano-rods: biomedical prospects} 
\author{Sayantan Ghosh$^*$, Jalpa Soni$^*$, Sudipta K. Bera, Ayan Banerjee and Nirmalya Ghosh
\skiplinehalf
Dept. of Physical Sciences, Indian Institute of Science Education and Research Kolkata, Mohanpur Campus, Mohanpur 741 252, Dist. Nadia, West Bengal, India\skiplinehalf}
\authorinfo{Further author information: (Send correspondence to S.G.); S.G.: E-mail: sg98@st-andrews.ac.uk; $^*$These authors contributed equally to the work}

\begin{document} 
  \maketitle 

\begin{abstract}
Fundamental understanding of the light-matter interaction in the context of nano-particles is immensely benefited by the study of geometry dependent tunable Localized Surface Plasmon Resonance (LSPR) and has been demonstrated to have potential applications in various areas of science. The polarization characteristics of LSPR in addition to spectroscopic tuning can be suitably exploited in such systems as contrast enhancement mechanisms and control parameters. Such polarization characteristics like diattenuation and retardance have been studied here using a novel combination of Muller-matrix polarimetry with the T-matrix matrix approach for silver nano-rods to show unprecedented control and sensitivity to local refractive index variations. The study carried out over various aspect ratios for a constant equal volume sphere radius shows the presence of longitudinal
(dipolar and quadrupolar) and transverse (dipolar) resonances; arising due to differential contribution of polarizabilities in two directions. The overlap regions of these resonances and the resonances themselves exhibit enhanced retardance and diattenuation respectively. The spectral and amplitude tunability of these polarimetric parameters through the aspect ratios to span from the minimum to maximum ($[0,1]$ in the case of diattenuation and $[0,\pi]$ in the case of retardance) presents a novel result that could be used to tailor systems for study of biological media. On the other hand, the high sensitivity of diattenuation dip (caused by equal contribution of polarizabilities) could be possibly used for medium characterization and bio-sensing or bio-imaging studies.
\end{abstract}
\section{INTRODUCTION}
\label{sec:intro}  
Localized Surface Plasmon Resonances (LSPR) due to their geometry dependent tunability have been explored in the recent years in various contexts of biomedical applications like sensing, development of novel nano devices \cite{lal2007}, probing bio-molecular interactions and their medical diagnostic and therapeutic applications \cite{Haes:2004} just to name a few. Localized surface plasmon resonances while exhibiting potential in wide ranging applications have also been probed for understanding fundamental physical questions related to the interaction of light with nano-particles \cite{maier2007,maier:011101}. The localized field enhancement exhibited when resonance between the incident electromagnetic radiation and polarizability of the nano-structure is achieved; have been understood to stem from the intrinsic dependence of metal polarizability on the shape and size of the particle. The spectral position of such plasmonic resonances are also well known to have an inherent dependence on the ambient medium. Such properties of the LSPR have been exploited (as mentioned before) to tailor them for specific applications. \par
The LSPR though traditionally investigated spectroscopically have recently been observed to show novel polarization properties \cite{jsoni2012} which can be used as additional contrast mechanisms in the study of light matter interaction. The effectiveness of such polarization characteristics (for example) in bio-sensing applications can be understood to arise from the fact that the elastically scattered spectra from dielectric components in complex biological media generally can swamp the plasmonic spectral signatures. However, it must be noted that polarization parameters like diattenuation and retardance are spectrally independent for dielectric structures while enhanced for plasmonic structures. This realization makes the study of polarization characteristics of plasmon resonance of considerable interest in the context of possible applications in various areas of sensing. 
\par
Plasmon resonant silver nano-rods have been investigated in the present context of biomedical applications recently owing to their interesting LSPR properties, where, polarization characteristics like diattenuation ($d$, defined as the differential attenuation between two
orthogonal polarizations), retardance ($\delta$, the phase difference between two orthogonal polarizations) and depolarization ($\Delta$, the randomization of the polarized light after passing through a medium); of silver nano-rods are probed for different aspect ratios, $\varepsilon\equiv d/l$ (where $d$ and $l$ are the diameter and length of the cylinder respectively) and ambient dielectric media. The T-matrix\cite{Mishchenko:96} method for calculation of the Mueller matrices and a decomposition algorithm based on Lu and Chipman's method have been used as analytical and computational tools in this study.
\par
This paper is organized as follows: section \ref{sec:theory} describes briefly the theoretical aspects of the work with a brief introduction to localized surface plasmons and the various analytical and computational tools used to study them. The extraction of the scattering matrices from the T-matrix method and their subsequent Mueller polarimetric study through a Lu and Chipman based approach is also described. Section \ref{sec:discuss} discusses the observations of the spectral characteristics of the LSPR for silver nano-rods along with the spectral  polarimetric parameters diattenuation, retardance and depolarization as a function of their geometrical configurations. The dependence of the polarimetric parameters on the ambient medium refractive index is also investigated and a quantification of the sensitivity of diattenuation to the medium refractive index is provided. Finally a conclusion and possible future directions is discussed in section \ref{sec:conc}.
\section{Theoretical aspects}
\label{sec:theory}
Surface plasmons can typically be described as collective oscillations of polarization charges generated at metal-dielectric interfaces, due to resonant illumination by electromagnetic waves and consequent conduction electron transport at the interface\cite{MaierLpr2012,maier2007,raether1988,vollmer1995}. The surface plasmons have been categorized as propagating (along the metal-dielectric interface) or localized (in sub-wavelength metal nano-particles). These resonances, owing to the polarization response of the metal nano-particles are highly sensitive to the geometry, shape, size and ambient medium. The shape and size dependence of the resonances can be easily understood if one considers that the polarizability of a sub-wavelength metal nano-particle is a function of its volume and consequently shape and aspect ratio ($\varepsilon$), which is the ratio of the lengths of the short axis and the long axis (e.g. for spheres, $\varepsilon=1$).
\par
Several methods have been developed for the analytical (where tractable) and numerical calculations of electromagnetic scattering from small particles. It is worthwhile to note here that the scattering problem is analytically quite easy to deal with in case of rotationally symmetric particles like spheres, spheroids and ellipsoids. Notable among such methods are the Quasi-static or Electrostatic Approximation (ESA)\cite{Bohren:2007,raether1988}, Mie theory\cite{mie:1908}, T-matrix method\cite{waterman1965,waterman1971}, Finite-Difference Time-Domain (FDTD) and Finite Element Method (FEM)\cite{sarid2010modern}.
\par
In the electrostatic approximation, the sub-wavelength ($a<<\lambda$) size of the particle allows the approximation that the particle lies in a static field and hence the polarizability can be easily calculated. For metal nano-spheres, the dipolar polarizability $\alpha(\omega)$ and the scattering cross-section $\csa$ can be calculated as\cite{mishchenkobook}
\begin{equation}
\alpha(\omega)=4\pi a^3\frac{\veq_1(\omega)-\veq_m(\omega)}{\veq_1(\omega)+2\veq_m(\omega)}\quad \csa=\pi a^2 \frac{8}{3}x^4\left\vert\frac{\veq_1(\omega)-\veq_m(\omega)}{\veq_1(\omega)+2\veq_m(\omega)}\right\vert^2.
\end{equation}
Here, $x=2\pi a/\lambda$ is the size parameter. It can easily be seen that for $\veq_1(\omega)=-2\veq_m(\omega)$ is singularity point where, the $\csa$ will sow significant enhancement\cite{Bohren:2007}. This is the dipolar resonance condition for the dipolar sub-wavelength metal nano-sphere. Similarly, the condition for quadrupolar resonance is given $\veq_1(\omega)=-(3/2)\veq_m(\omega)$. In general, for spheres, spheroids and ellipsoids, the dipolar polarizability given by
\begin{equation}
\alpha_j(\omega)=4\pi a b c\frac{\veq_1(\omega)-\veq_m(\omega)}{3\veq_m(\omega)+3L_j(\veq_1(\omega)-\veq_m(\omega))};
\end{equation}
where, for spheres, $a=b=c$ which implies that $L_1=L_2=L_3=1/3$, for spheroids, $a=b\neq c$ and for ellipsoids, $a\neq b\neq c$. 
\par
Although sub-wavelength particles can be easily studied using the quasi-static approximation, this method fails when subjected to particles whose sizes are comparable to that of the wavelength ($r\lesssim\lambda$). Gustav Mie in his pioneering work proposed the expansion of the incident and scattered electric fields in vector-spherical wave functions which could be used to solve the Maxwell's equations along with the boundary conditions (continuity of the electric and magnetic fields at the particle medium interface) exactly albeit only for spherical particles\cite{mie:1908,Bohren:2007}. The solution of the Maxwell's equations in this case take the form of spherical Bessel and Hankel functions. This method allows us to write the relation between the incident and scattered electric field components in the Jones formalism by $\vec{E}_{s}=\mathbf{J}\vec{E}_{i}$ where, the $E_\parallel$ and $E_\perp$ are parallel and perpendicular to the scattering plane respectively and $\mathbf{J}$ is the $2\times 2$ amplitude scattering matrix or the Jones matrix which has a diagonal form for spheres. The elements of the amplitude scattering matrix $\mathbf{J}$ are represented by the scattering co-efficients $a_n$ nd $b_n$ obtained from the Mie theory calculations\cite{Bohren:2007}. The index $n$ in the scattering coefficients represents the multipolar modes (e.g. $n=1,2$ correspond to dipolar and quadrupolar modes respectively) and for each $n$, the corresponding $a_n$ and $b_n$ can be interpreted as the transverse magnetic or TM (no radial magnetic field) and transverse electric or TE (no radial electric field) modes respectively\cite{Bohren:2007}. Since the Mie theory is applicable only for spheres, various extensions of this method to non-spherical particles like ellipsoids and cylinders have been proposed. However, the most prominent amongst them is the T-matrix method which calculates the scattering matrix for arbitrarily shaped particles by solving the Maxwell's equations while expanding the incident and scattered electric fields in terms of the vector spherical wave harmonics\cite{waterman1965,waterman1971} and is computationally best suited to the calculation for rotationally symmetric particles\cite{Mishchenko:96,mishchenkobook}. This T-matrix method analogous to the Mie theory yields the relation between the incident and the scattered electric fields's Stokes vectors ($\vec{S}_i$ and $\vec{S}_s$ respectively) through a $4\times 4$ matrix called the scattering matrix or the Mueller matrix (named after Mueller for the Stokes-Mueller algebra), such that, $\vec{S}_s=\mathbf{M}\vec{S}_i$\cite{Bohren:2007,Mishchenko:96}. The Mueller matrix by the virtue of its formulation contains all the polarization information about the scattering medium and the particle\cite{Broseau:1998} and can be explored using various Stokes-Mueller algebra based methods to extract and quantify polarization properties of scattering\cite{NGhosh:2008,NGhosh:2010,NGhosh:2011}. The polarization parameters thus obtained have been very important both in furthering the basic understanding of light-matter interaction\cite{Ronchi:1970,Broseau:1998} and their applications to various front-line research areas like astronomy\cite{Lucas:2007}, meteorology\cite{Gadsden:1979}, and biomedical research\cite{NGhosh:2008, NGhosh:2010,NGhosh:2011}. 
\par
The study of the polarization characteristics of light matter interaction requires careful consideration of techniques that would provide the maximum information about the systems under study. For example, the Jones polarimetry involves coherent addition of amplitudes and phases of the electric field and is experimentally difficult to carry out. This technique also fails to account for the effects of partial polarization aspects\cite{Broseau:1998}. The Stokes-Mueller based technique is an intensity based technique and has emerged to be one of the most extensive polarimetric tools available to the community. As mentioned earlier, the incident and the scattered electric fields can be related by a $4\times 4$ transfer matrix or Mueller matrix $\textbf{M}$ which carries polarization information in both linear and circular bases. The elements of Mueller matrix $\textbf{M}$ carry information like linear and circular depolarization $(\Delta_L \& \Delta_C)$, retardance $(\delta_L \& \delta_C)$ and diattenuation $(d_L \& d_C)$ respectively. Depolarization is defined as the decrease in the net degree of polarization (which is defined as $\left.\textrm{DOP}=\frac{\sqrt{Q^2+U^2+V^2}}{I}\right)$ due to the sample. Depolarization is given as $\Delta=1-\textrm{DOP}$, where the incident EM-field is assumed to have a net $\textrm{DOP}=1$. Depolarization can be understood to originate from multiple scattering processes in the sample. Retardance $\delta$ is defined as the phase shift between two orthogonal polarization states. This property arises from the existence of polarization axes in axisymmetric particles like ellipsoids and cylinders. This phase difference for linear orthogonal polarization states like $(H,V)$ and $(P,M)$ is called the linear retardance $\delta_L$ and for circular orthogonal polarization states like $(L,R)$ is called as circular retardance $\delta_C$. Similarly, diattenuation $d$ is defined as the differential attenuation of orthogonal polarization states while interacting with a sample and has the same origin as retardance. The differential attenuation of orthogonal linear (like ($H,V$) and ($P,M$)) and circular (like $(L,R)$) polarization states give linear $(d_L)$ and circular $(d_C)$ diattenuation respectively. 
\par
Quantification and extraction of ``lumped'' polarization properties requires a mathematical decomposition of the Mueller matrix $\textbf{M}$, which can be done in various ways. One such method is the Polar Decomposition Method (PDM) following the prescription of Lu and Chipman\cite{Lu:96}. The $4\times 4$ Mueller matrix $\textbf{M}$ can be decomposed into three ``basis'' matrices based on its polarization properties. This polar decomposition method sequentially extracts diattenuating $\mathbf{M}_d$, retarding $\mathbf{M}_R$ and depolarization $\mathbf{M}_\Delta$ matrices as
\begin{equation}
\mathbf{M}\Leftarrow \mathbf{M}_\Delta \cdot \mathbf{M}_R \cdot \mathbf{M}_D,
\end{equation}
where $\Leftarrow$ represents the decomposition where, the decomposed basis matrices are defined as
\begin{equation}
\mathbf{M}_D=\begin{pmatrix}1& \vec{d}^T\\\vec{d}&m_d\end{pmatrix},\mathbf{M}_\Delta=\begin{pmatrix}1 &\vec{0}^T\\P_\Delta &m_\delta\end{pmatrix},\mathbf{M}_R=\begin{pmatrix}1&\vec{0}^T\\\vec{0}&m_R\end{pmatrix}.\nonumber\end{equation} Here, $\vec{d}=\frac{1}{M_{11}}[M_{12}~M_{13}~M_{14}]^T$ is the diattenuation vector, $\hat{d}=\vec{d}/\vert\vec{d}\vert$ is the unit vector and $m_D=\sqrt{1-d^2}\mathbf{I}+(1-\sqrt{1-d^2})\hat{d}\hat{d}^T$. The magnitude of the diattenuation vector $\vert\vec{d}\vert$ gives the total diattenuation of the system and is expressed as $d=\frac{1}{M_{11}}\sqrt{M_{12}^2+M_{13}^2+M_{14}^2}$. Similarly, $P_\delta=\frac{\vec{P}-m\vec{d}}{1-d^2}$, where, $\vec{P}=\frac{1}{M_{11}}[M_{21}~M_{31}~M_{41}]^T$ is the polarizance vector. $m_\delta$ and $m_R$ are $3\times 3$ sub-matrices of $\mathbf{M}_\Delta$ and $\mathbf{M}_R$ respectively and can be found by solving the eigenvalues of $m'=m_\delta\cdot m_R$. The net depolarization can be calculated as $\Delta=1-\frac{\vert \textrm{tr}(M_\Delta)-1\vert}{3}, 0\leq \Delta \leq 1$ while the net retardance ($R$) is given by $R=\cos^{-1}\left(\frac{\textrm{tr}(M_R)}{2}-1\right)$. The linear retardance can be calculated from the $\mathbf{M}_R$ as $\delta=\cos^{-1}\left[\sqrt{(M_{R,22}+M_{R,33})^2+(M_{R,32}+M_{R,23})^2}-1\right]$.
\section{Results and discussion}
\label{sec:discuss}
Silver nano-particles and nano-rods in particular have attracted the attention of the scientific community in the recent years in context to their varied applications. In this work, we consider silver nano-rods (circular cylinders) of equal-volume-sphere-radius (EVSR) $20$nm and varying aspect ratios. Unless mentioned otherwise, the EVSR will be considered to be $20$nm and the propagation direction of the incident electromagnetic field will be positive $Z$ direction throughout the work. The refractive index data of Winsemius \textit{et al.} \cite{dataref} has been used for the calculations. As has been mentioned before, the plasmonic signature of the metal nano-particle is affected by its orientation and the ratio of its short axis $2r$ ($r$ being the radius) to its long axis $l$ ($l$ being the length of the nano-rod). Hence, we investigate the electromagnetic scattering and the polarization response of these silver nano-rods as function of their aspect ratio $\varepsilon=2r/l$. The detection plane is again the $X-Z$ plane with the detection angle $\theta=0^\circ$ signifying ``on-axis'' detection, i.e. the detection point lies on the propagation axis. $\theta=90^\circ$ detects the scattering at an angle perpendicular to the propagation direction; this can be either on the $\pm X$ axes. In this study, we have chosen the scattering angle $\theta=45^\circ$. The scattering cross section $\csa$ and the Mueller matrices have been obtained using the T-matrix codes written in Fortran77 by Mishchenko \textit{et al.}\cite{Mishchenko:96,Mishchenko:00,mishchenkobook} 
\par
The silver nano-rods by the virtue of a cylindrical geometry can be understood to have two polarization axes along the long and short axes of cylinder. For example, when the particle's long axis is aligned along the propagation direction of the incident radiation, a transverse axis is formed in the $X-Y$ plane and a longitudinal axis is formed along the long axis of the particle. The particle size also plays a crucial role in the excitation of the plasmon modes. It is well known that due to the nature of the permittivity response of silver to electromagnetic radiation, the excitation of different orders of plasmon modes is limited. In fact, for spheres the minimum limit for quadrupolar excitation is $50$nm\cite{maier:011101}. In rods, however, at an equal-volume-sphere-radius of $20$nm, the aspect ratio $0.65$ corresponds to a rod length of approximately $44.9$nm and a rod diameter of $30$nm, it is possible to excite quadrupolar resonances, albeit only in the transverse mode. The absence of the longitudinal quadrupolar mode can be understood easily from the consideration that the alternating positive-negative charge separations along the length of the cylinder cannot be sustained over such lengths. However, the longitudinal dipolar mode can be sustained over the length of the cylinder.
\begin{figure}[h]
\centering
\includegraphics[clip,bb= 31 182 563 615,width=0.45\textwidth]{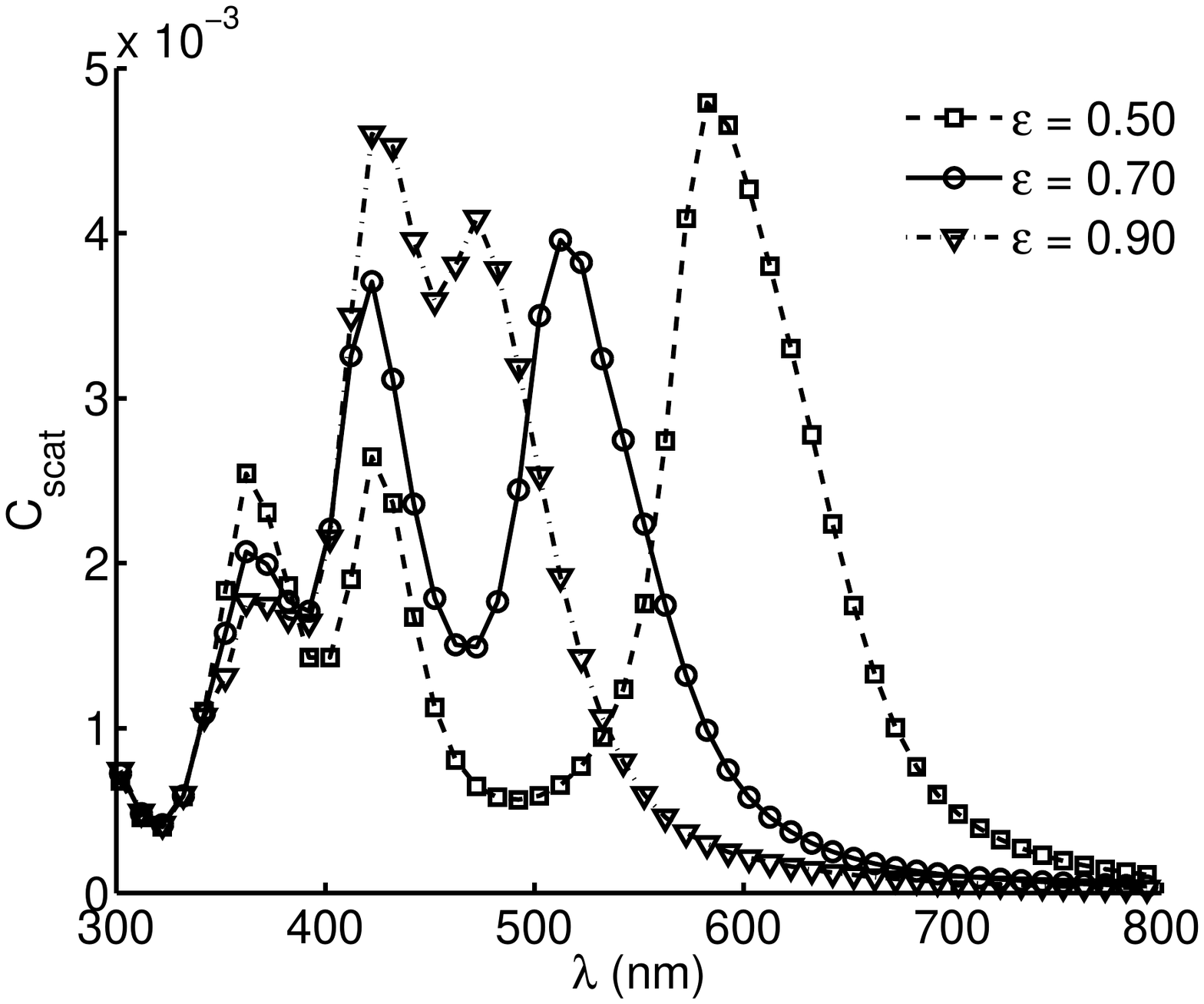}
\caption{\label{fig:figscat}The spectral scattering cross section $C_\textrm{scat}$ for varying aspect ratios $\varepsilon$ exhibiting the quadrupolar and dipolar resonances.}
\end{figure}
\par
Figure \ref{fig:figscat} shows the orientation averaged spectral variation of the extinction cross-section for nano-rods of EVSR $20$nm at different aspect ratios. As expected, the quadrupolar plasmon mode is excited at wavelengths lower than the dipolar modes. The quadrupolar mode (excited at $\sim375-400$nm) is transverse in nature and thus remains unaffected by the change in the aspect ratio which entails changes in the length of the nano-rods. The two dipolar modes, one transverse and the other longitudinal are excited in the spectral range $\sim410-440$nm and $\sim 600$nm (only for $\varepsilon=0.50$) respectively. The longitudinal dipolar resonance is observed to blue shift with increasing aspect ratio. As the aspect ratio approaches $\varepsilon\approx1$, the two transverse and longitudinal dipolar resonances merge. This can be attributed to the equal contribution of the respective polarizabilities. The relative intensities of the plasmon modes are also controlled by the strength of relative polarizability contributions. We must note here that the behavior of the resonances for nano-rods is a departure from that for ellipsoids where the transverse and longitudinal resonances get blue and red shifter respectively with increasing aspect ratio\cite{jsoni2012}. 
\par
As mentioned earlier, the polarization characteristics described the polarimetric parameters like depolarization ($\Delta$), diattenuation ($d$) and retardance $(\delta)$ not only provide a wealth of information about the system under consideration, but like in plasmonic resonances, can be used as additional contrast parameters for high precision sensing applications. In the following, we investigate the spectral response of the polarimetric parameters in the context of their dependence on the aspect ratio. The calculations have been undertaken by preferentially orienting the long axis of the particle with the propagation direction of the incident radiation. The Mueller matrices obtained using the T-matrix method for preferentially oriented particles were subjected to the decomposition method described in the earlier section.
\begin{figure}[h]
\centering
\subfigure[$d$ at $\theta=50^\circ$]{\includegraphics[clip,bb= 31 182 563 615,width=0.31\textwidth]{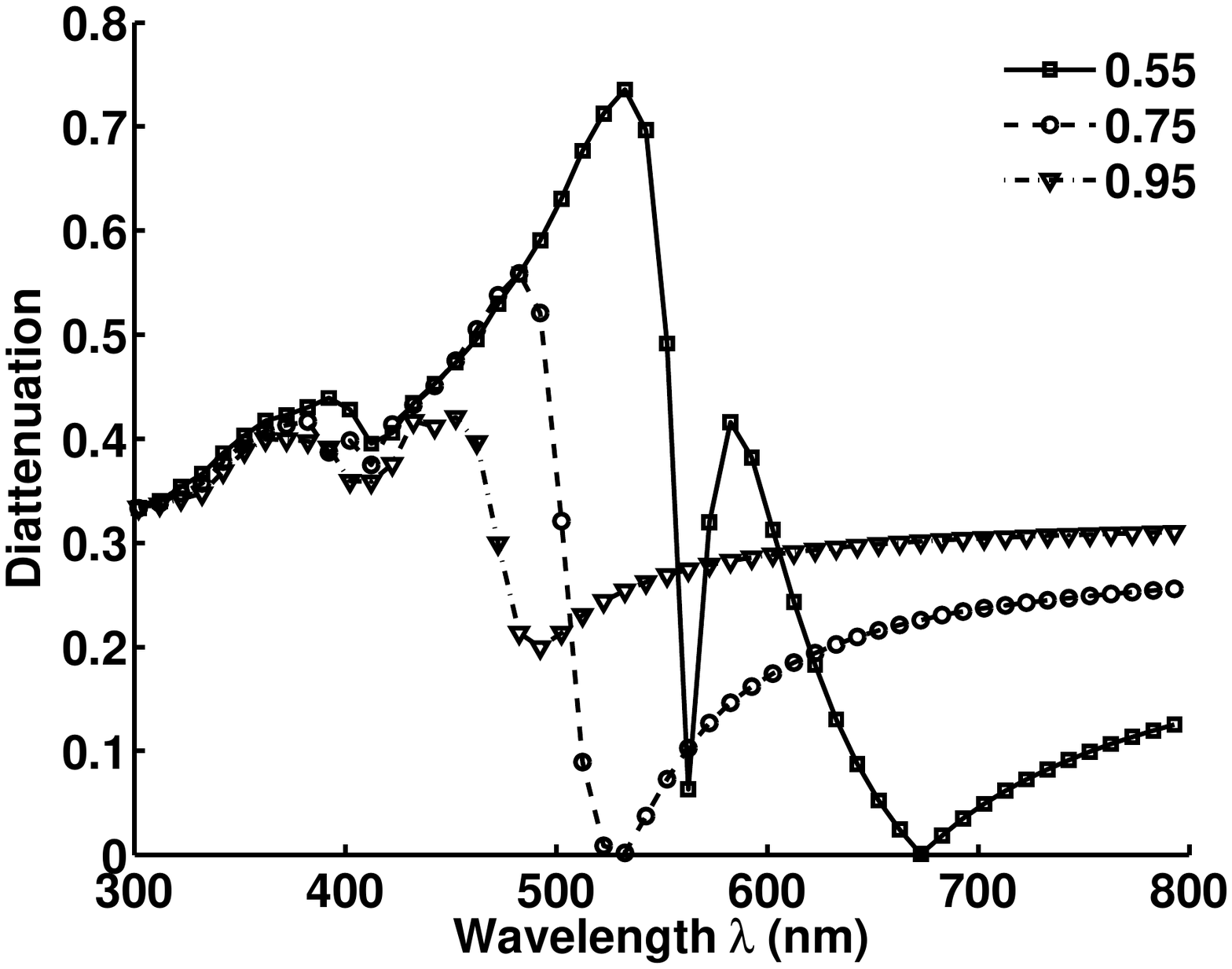}\label{fig:fig4a}}
\subfigure[$\delta_L$ at $\theta=50^\circ$]{\includegraphics[clip,bb= 31 182 563 615,width=0.31\textwidth]{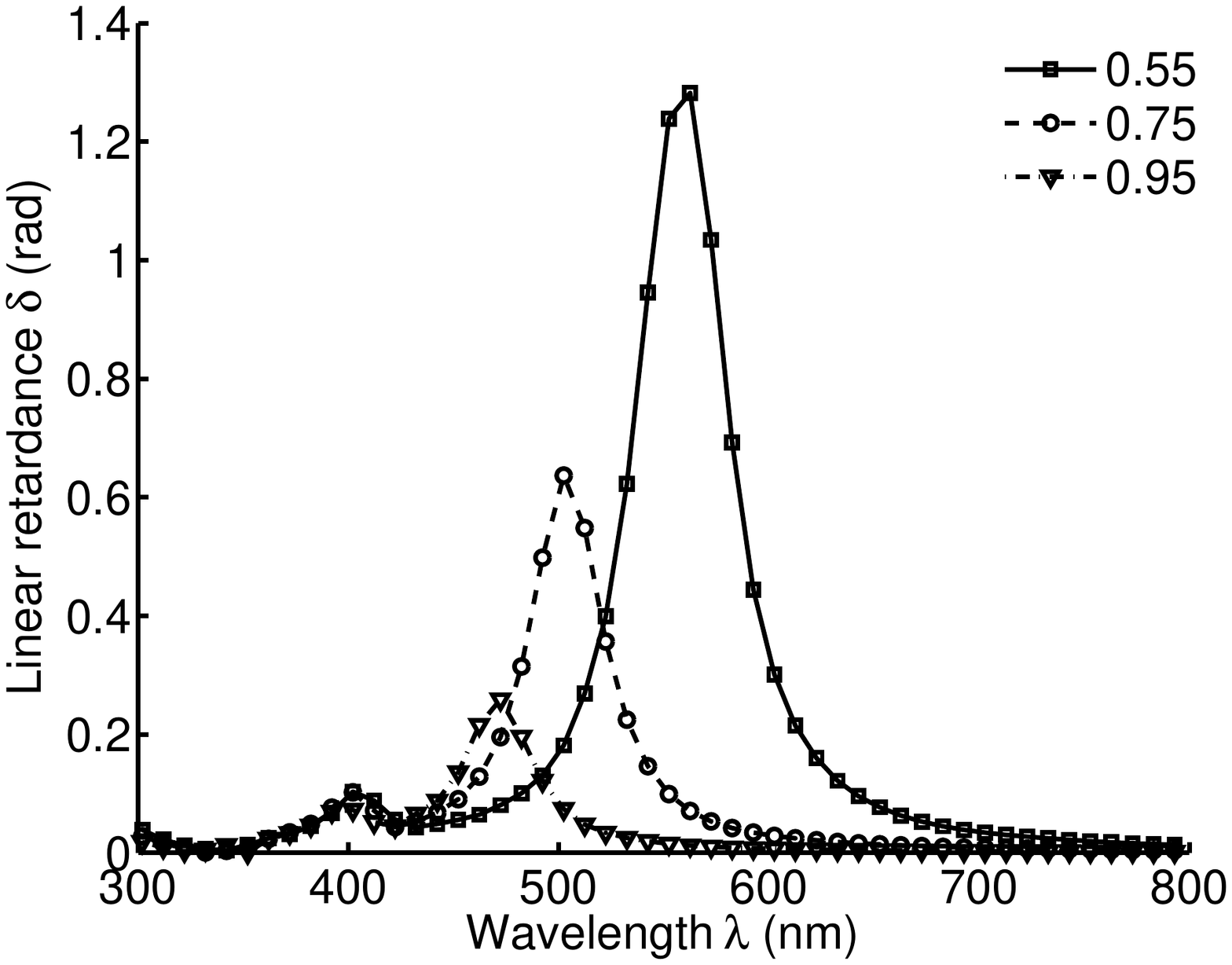}\label{fig:fig4b}}
\subfigure[$\Delta$ at random orientation]{\includegraphics[clip,bb= 31 182 563 615,width=0.31\textwidth]{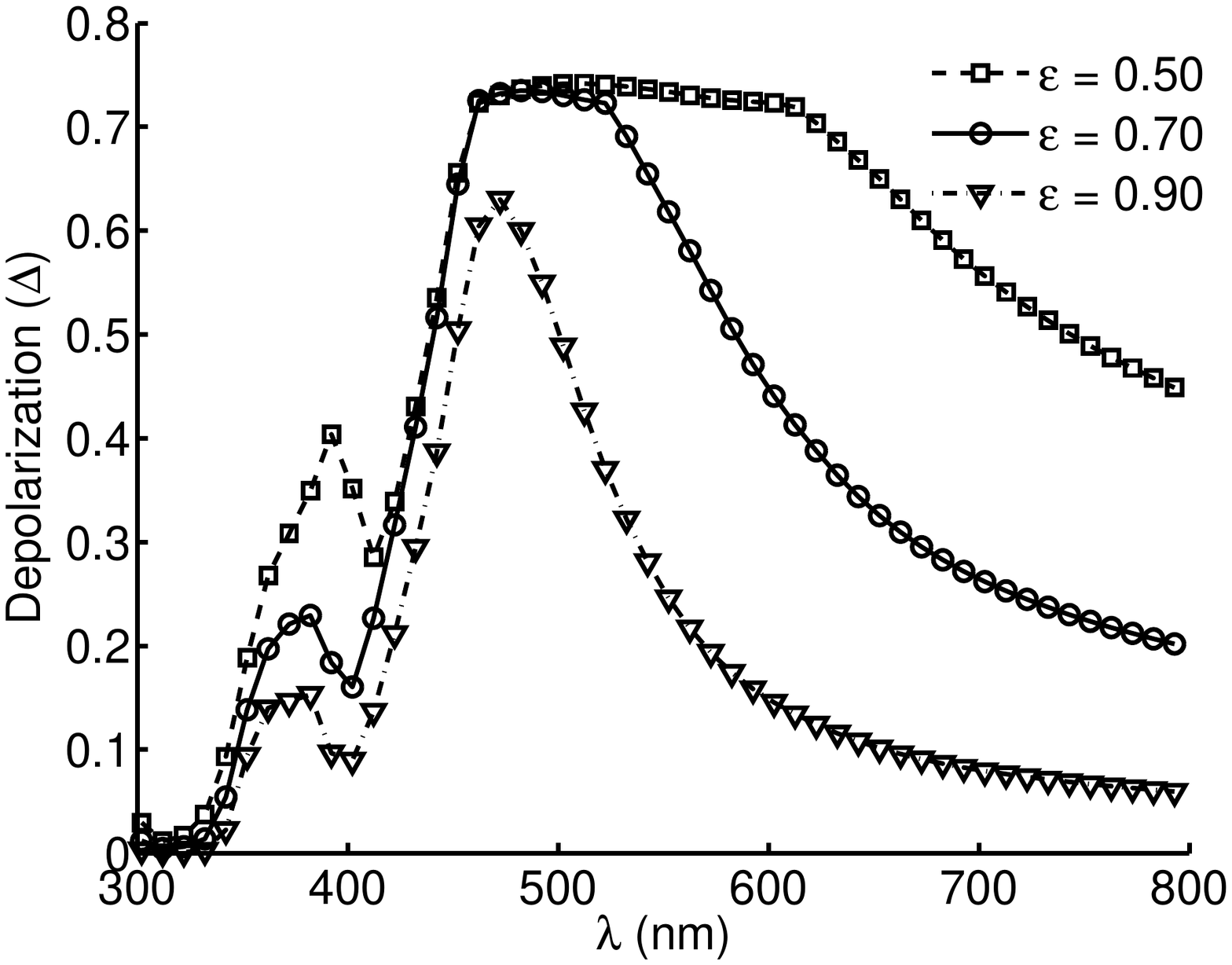}\label{fig:fig4c}}
\caption{\label{fig:fig4}\subref{fig:fig4a}Linear retardance $\delta_L$ and \subref{fig:fig4b} diattenuation $d$ with different aspect ratios for the nano-rod oriented at a zenith angle $0^\circ$ to the laboratory frame of reference. \subref{fig:fig4c} shows the net depolarization for orientation averaged, randomly oriented nano-rods. The scattering angle $\theta$ in each case was kept to be $50^\circ$}
\end{figure}
\par
Linear diattenuation and retardance as mentioned earlier are the differential attenuation and phase difference between linear orthogonal polarizations. The linear diattenuation and retardance for silver nano-rods of EVSR $20$nm whose long axis is oriented along the propagation direction of the incident field, are shown in Figures \ref{fig:fig4a} and \subref{fig:fig4b}. In \ref{fig:fig4a}, the linear diattenuation can be seen to peak at the plasmonic resonances while at the spectral overlap regions of two competing transverse and longitudinal modes is observed to show a sharp concavity or dip. However, it can be seen that as the aspect ratio $\varepsilon$ increases, the dip becomes broader. This can be attributed to the equal contributions of the transverse and longitudinal modes when the rod approaches a disk. In contrast to the diattenuation, the linear retardance spectrum can be seen to peak in regions where the plasmon modes overlap. Nano-rods as has been mentioned earlier have two polarization axes effecting the polarizability to have two orthogonal components. The preferential orientation of the particle at $\beta=0^\circ$ aligns the two polarization axes to coincide with the orthogonal polarizations of the incident electromagnetic radiation. This causes a differential attenuation of the two orthogonal polarization states while interacting with the metal nano-particle, the effect of which can be seen as higher diattenuation at the resonance bands. At the spectral overlap regions of the two bands, the polarizability can be understood to be the vector sum of two orthogonal polarizations which would be commensurate with the low diattenuation as shown in Figure \ref{fig:fig4a}. However, at these spectral overlap regions, the phase difference between the two orthogonal polarizabilities is high which shows up as the increased linear retardance.
\par
Depolarization as mentioned before arises from multiple random scattering in the sample and as mentioned earlier is represented as the loss of polarization for 100\% polarized incident electromagnetic radiation. In Figure \ref{fig:fig4c}, the depolarization spectrum is shown. The fact that the spectral signature of the depolarization closely resemble that of the linear retardance spectrum can be understood as the effect of the agglomeration of multiple retarders in random orientations giving rise to depolarizations. In fact, when the particle is preferentially oriented, the depolarization matrix $M_\Delta=M_R$.
\par
The diattenuation described in figure \ref{fig:fig4a} was shown to exhibit a very sharp concavity at the spectral overlap region of the two competing transverse and longitudinal resonances. This sharp concavity as has been mentioned before is more prominent for aspect ratios much lower than $1$ and the the concavity becomes spectrally wider as the aspect ratio approaches $1$. This dip as we shall show presently, can be possibly used as a bio-sensor that can detect small fluctuations in the medium refractive index. 
\begin{figure}[h]
\centering
\subfigure[Diattenuation with medium refractive index]{\includegraphics[clip,bb= 31 182 563 615,width=0.45\textwidth]{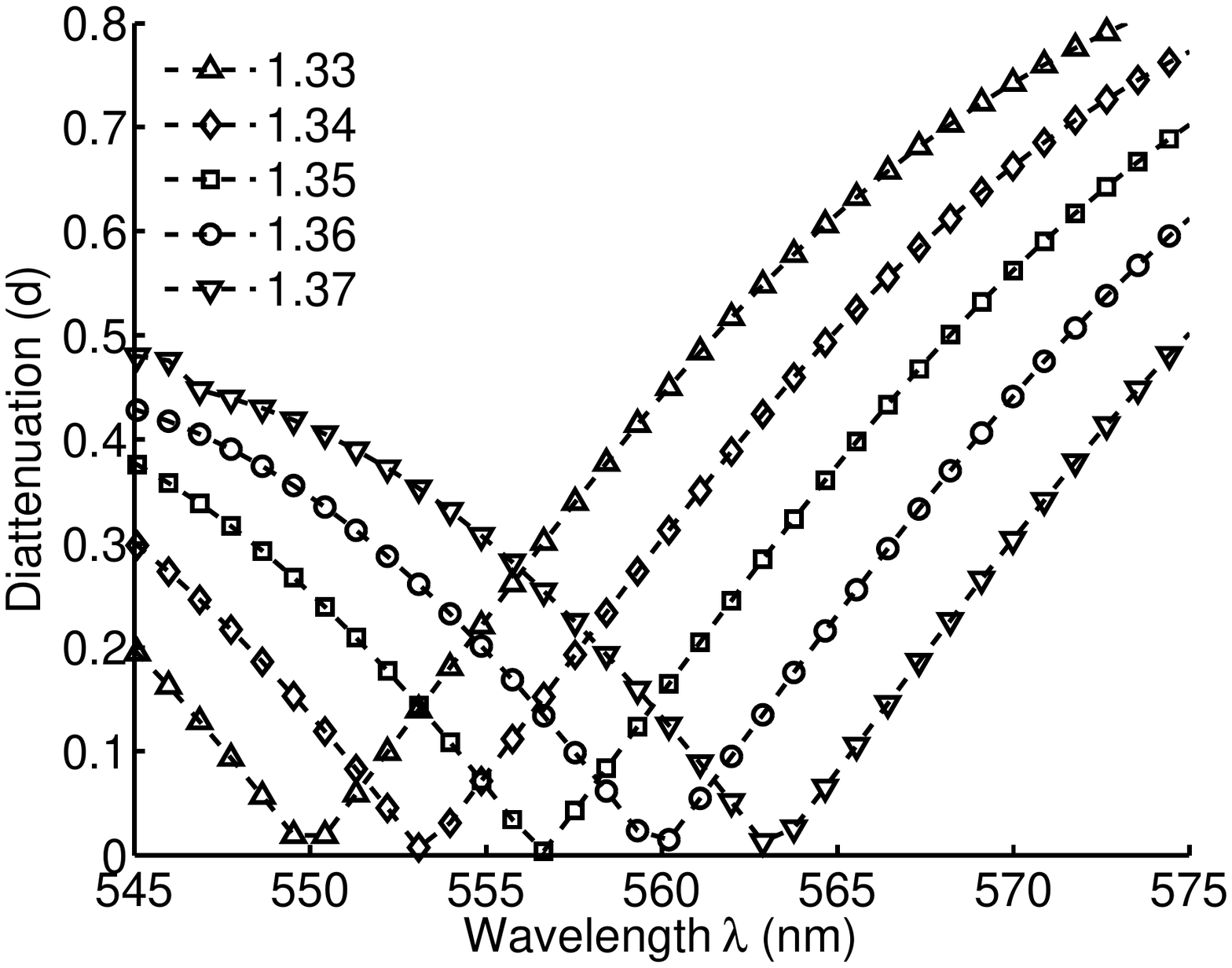}\label{fig:figri}}
\subfigure[figure of merit $\gamma$]{\includegraphics[clip,bb= 31 182 563 615,width=0.45\textwidth]{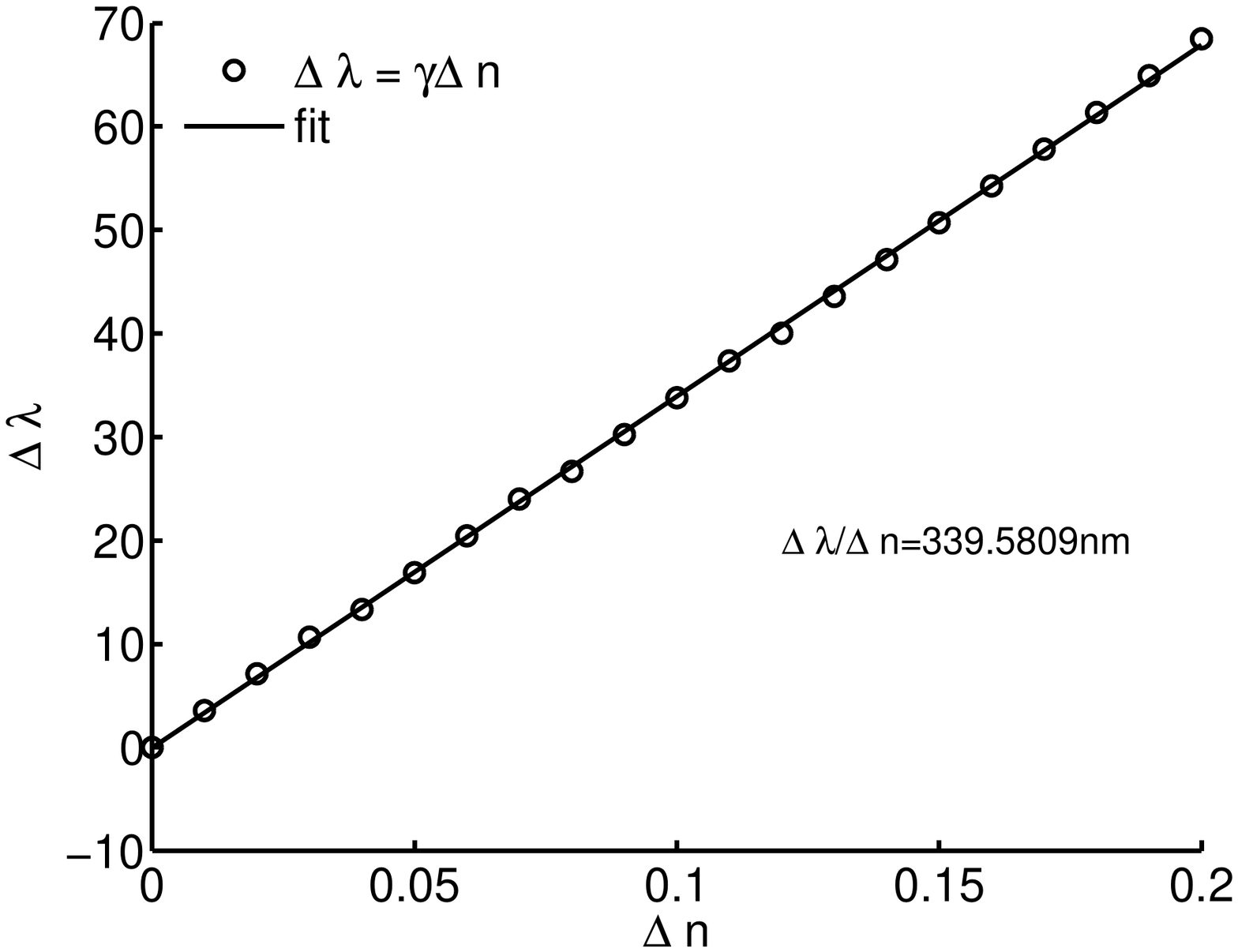}\label{fig:figf}}
\caption{\label{fig:figfom} Here in \subref{fig:figri} Sensitivity of the diattenuation-dip of a nano-rod ($\varepsilon=0.65$, oriented at an azimuthal angle $30^\circ$ with the laboratory frame) as a function of wavelength ($\lambda$) is shown for different values of surrounding medium refractive index and \subref{fig:figf} shows the figure of merit $\gamma$ a representative case.}
\end{figure}
Figure \ref{fig:figri} represents the spectral sensitivity of the diattenuation dip of a silver nano-rod ($\varepsilon=0.65$, oriented at an azimuthal angle $30^\circ$ with the laboratory frame) with refractive index of the medium. It can clearly be seen that as the medium refractive index $n_m$ increases, the diattenuation dip red shifts by $\sim 3-4nm$ with every $10^{-2}$ increase in the $n_m$. This spectral sensitivity can be understood while seen in conjunction with the scattering cross-section where, the spectral position of the longitudinal dipolar resonance also red shifts with increase in $n_m$. The figure of merit $(\gamma)$ for the variation in spectral position of the diattenuation dip with varying medium refractive indices has been calculated as
\begin{equation} 
\gamma=\frac{\Delta\lambda}{\Delta n}
\end{equation}
and has been shown in figure \ref{fig:figf}. The figure of merit for the representative case shown in figure \ref{fig:figri} is found to be $339.5809nm$ which signifies a spectral shift of $339.5809nm$ for the diattenuation dip for unit change in the medium refractive index. 
\par
The fact that the diattenuation for plasmonic nano-particles is significantly higher than that of dielectric nano-particles of similar dimensions, and the spectral dependence of such diattenuation for metal nano-particles (arising from the spectral dependence of metal polarizabilities) can be exploited as a contrast mechanism for the detection of local refractive index variations in various inhomogeneous media where, the homogeneous volume is significantly higher than the particle's volume. This diattenuation dip could consequently be used in various biomedical applications like real-time sensing of a biological cell's growth, determination of the sub-micron refractive index structure of tissues and provide various tools which could possibly used to probe disease progressions.
\section{Conclusion}
\label{sec:conc}
To summarize, characterization of LSPR in plasmon resonant silver nano-rods was carried out in this work using polar decomposition of scattering Mueller matrices in different configurations. The presence of quadrupolar resonance along with the dipolar resonance in the electromagnetic scattering spectra of such nano-rods show interesting effects on the spectral features of the polarimetric parameters. The decomposition analysis from the preferentially oriented particles revealed the presence of strong diattenuation ($d$) and linear retardance ($\delta$) effects in the regions corresponding to the resonances and in the overlapping regions of the resonances respectively. Distinct spectral features for both diattenuation and retardance were observed and are found to be sensitive to the size and orientation of the nano-rods. These polarimetric parameters were further investigated for the nano-rods suspended in various dielectric media having different refractive indices and it was observed that small change in the medium refractive index leads to dramatic changes in the spectral response of both the retardance and diattenuation. However, the dip of the diattenuation arising between transverse and longitudinal dipolar modes was found to be maximally sensitive to the medium refractive index variations, with the figure of merit being $339.5809 nm$. This high sensitivity of the diattenuation dip, which experiences a spectral shift of more than $300 nm$ with a unit change in the medium refractive index has been further probed with an emphasis on biomedical applications. We have used the values of the medium refractive index corresponding to the local refractive index variations in biological systems and observed that the diattenuation dip does show sensitivity to such small changes and hence can successfully be exploited and used as the contrast mechanism in various biological applications. It is also worth noting that the unprecedented control over the polarimetric parameters for these silver nano-rods could possibly also be gainfully explotied for addressing fundamental questions in light-matter interaction, the details of which are under current investigation.
\section*{Acknowledgements}     
S.G. would like to thank IISER-Kolkata for hosting and funding this internship and Subimal Deb for valuable discussions.

\end{document}